\documentclass[apj]{emulateapj}

\usepackage{amssymb}
\usepackage{epsfig}
\usepackage{graphicx}

\begin{document}

\title{What Millisecond Pulsars Can Tell Us About Matter In The Galaxy}

\author{E. R. Siegel$^{1,2}$}
\affil{$^{1}$Steward Observatory, University of Arizona, 933 N. Cherry
Avenue, Tucson, AZ 85721 \\
$^{2}$Department of Physics, University of Wisconsin, 1150
University Avenue, Madison, WI 53706}


\begin{abstract}
I demonstrate that precision timing of millisecond pulsars possess the capabilities of detecting the gravitational effects of intervening galactic substructure.  This analysis is applicable to all types of collapsed baryons including stars, planets, and MACHOs, as well as many types of dark matter, including primordial black holes, scalar miniclusters, and sufficiently dense clumps of cold dark matter.  The physical signal is quantified and decomposed into observable and unobservable components; templates for the observable signals are also presented.  Additionally, I calculate the expected changes in the observed period and period derivatives that will result from intervening matter.  I find that pulsar timing is potentially a very useful tool for probing the nature of dark matter and to learn more about the substructure present within our galaxy.
\end{abstract}


\section{Introduction}

Our modern understanding of cosmology has revolutionized our picture of our own galaxy.  What was once thought to be an island in the universe dominated by stars is now better modeled as a disk of mostly non-luminous baryons embedded in a much more massive halo composed primarily of dark matter \citep{Cole:2005sx}.  However, the mass distribution and clumpiness of the both the disk and the halo, as well as the nature of the dark matter, are at present unknown.

Baryons collapse to form molecular clouds, stars, and planets, among other things \citep{FP:04}.  The location and abundance of the baryonic components of the galaxy are known primarily from their optical effects, such as emission and absorption features.  This tactic does not work for dark matter, though, since it does not interact electromagnetically with any appreciable strength \citep{CDMS:2005}.  However, both dark matter and baryons do interact gravitationally.  Thus, if a galactic probe that is highly sensitive to the gravitational influence of the components of the galaxy can be identified, it will not only have the potential to detect the baryons that optical searches have missed thus far, but can also be used to detect dark matter within our own galaxy.

Such a galactic probe highly sensitive to gravitational effects has been known since 1982: the millisecond pulsars \citep{MSP:82}.  Over one hundred millisecond pulsars have already been identified in our galaxy, both in the disk and the halo \citep{ATNF}.  Because of the stability and high accuracy of the timing measurements of millisecond pulsars, even small changes in the gravitational environment as the light from a pulsar travels to Earth can, in principle, be detected \citep{LD:95,Fargion:97,Hosokawa:1999,Siegel:2007fz,Seto:2007}.

With this paper, I seek to quantify the physical effects and observable signatures of baryonic and dark matter on the timing measurements of millisecond pulsars, with a view towards using millisecond pulsars to detect heretofore invisible baryons and dark matter.  The layout of this paper is as follows: section 2 discusses the physical gravitational effects of matter on the light-travel-time from a pulsar to Earth.  Section 3 focuses on what component of the physical signal is uniquely ascribable to this gravitational effect, identifying points of confusion and their resolution.  In section 4, templates for identifying transiting matter are created and presented, along with probabilities for detecting dark matter via this method.  Finally, section 5 concludes this paper with a discussion of possible applications and future work as well as briefly addressing more complicated models of dark matter.

\section{Gravitational Effects of Matter}

When considering the gravitational effect of matter on the light-travel-time of a signal emitted from a pulsar, one must consider the gravitational potential at every point along which the signal travels.  A constant gravitational potential will result in no changes in the light-travel-time, but temporal variations in the gravitational potential will lead to time variations due to the Shapiro time delay effect \citep{Shapiro:1964}.  This effect has been well-studied within our own solar system, as, for example, the moon causes periodic variations every twenty-eight days in the arrival time of pulses from all pulsars \citep{Will:80}.  It has also been used to discover planets and binary companions to certain pulsars \citep{Konacki:2003}.  In this work, I address the possibility that there will be changes in the gravitational potential along the line-of-sight (LOS) due to intervening matter.  Note that this phenomenon is physically similar to the microlensing phenomenon, where a clump of matter causes a predictable change in brightness of a background source.  Meaningful constraints have already been placed on the MACHO content of the galactic halo from microlensing limits \citep{Alcock:98}. However, I argue that pulsar timing has the potential to be an even more sensitive probe than microlensing of both collapsed baryons and, particularly, of non-baryonic dark matter.

\begin{figure}
\includegraphics[width=3.5in]{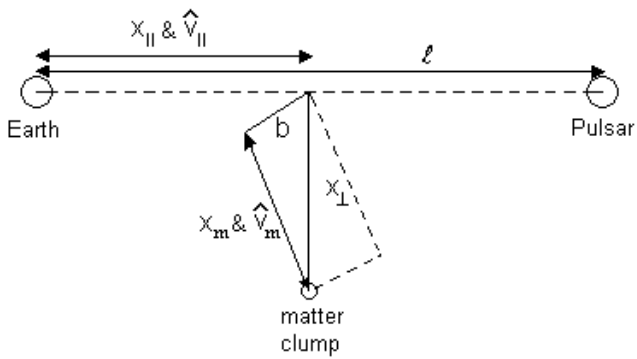}
\caption{Configuration for a transiting matter clump near the line-of-sight (LOS) from Earth to a pulsar located a distance $l$ from Earth.  Parallel components of position $(x_\parallel)$ and velocity $(v_\parallel)$ of the matter clump are along the LOS, while perpendicular components are in the plane perpendicular to the LOS.  The perpendicular component of velocity is defined to be in the direction $v_m$, while the direction perpendicular to both the LOS and $v_m$ is defined as the direction of the impact parameter, $b$.}
\label{config}
\end{figure}

I consider the following configuration as illustrated in figure \ref{config}, where a pulsar located at a fixed distance $l$ from Earth has a clump close to the LOS of (presumably non-luminous) matter of mass $M_\mathrm{dm}$.  The position of the massive clump is given by the components $x_\parallel$, which is the distance from Earth to the clump, and $\vec{x}_\perp$, which is the perpendicular distance of from the LOS to the clump.  $\vec{x}_\perp$ can be further broken up into components $\{x_m \mathrm{,} \, b\}$, where $b$ is the impact parameter, and $x_m$ is the component of position that will change over time due to the peculiar velocity $\vec{v}$ of the clump, which itself has components $v_\parallel$ and $v_m$.

Baryons make up approximately one-sixth of the matter in the galaxy, with dark matter composing the other five-sixths \citep{Wilk:1999}.  I estimate the total mass of the galaxy to be $\sim 1.2 \times 10^{12} \, M_\odot$, and define $f_\mathrm{dm}$ as the fraction of galactic matter (baryonic and dark) found in clumps of mass $M_\mathrm{dm}$.  Each clump is estimated to have a typical velocity of $|\vec{v}| \simeq 200 \, \mathrm{km} \, \mathrm{s}^{-1}$ relative to the LOS to the pulsar, evenly distributed among its components, $v_m$ and $v_\parallel$.  This is a reasonable estimate for the velocity relative to the LOS to the pulsar, as millisecond pulsars typically have large peculiar velocities relative to the bulk flow of nearby baryonic matter.

I assume henceforth that the clump of intervening matter is either point-like or sufficiently far away from the LOS that the impact parameter ($b$) is greater than the physical radius of the clump of matter.  Although this is not necessarily true for all forms of matter (WIMPy dark matter microhalos are the obvious exception), the calculations involved in computing the effects of extended dark matter microhalos are beyond the scope of this paper.  With the clump of intervening matter treated as effectively point-like, the Shapiro time delay $(\Delta t)$ due to any point-like mass $M_\mathrm{dm}$ is
\begin{equation}
\label{Tdelay} \Delta t = \frac{2 G M_\mathrm{dm}}{c^3}
\ln{\left(\frac{4 x_\parallel (l -
x_\parallel)}{x_\perp^2}\right)} \mathrm{,}
\end{equation}
where $x_\parallel$, $x_\perp$, and $l$ are defined above.  

Equation (\ref{Tdelay}) is the time delay induced by a clump of mass as compared with not having that mass present at all.  However, physically, the mass is always present, and merely changes its position over time.
In this scenario, the dark matter will transit from an initial
position at an initial time $(t_i)$, and thus cause an initial
time delay $(\Delta t_i)$, to a final position $(t_f)$ with a
final time delay $(\Delta t_f)$, leading to a physical delay
$(t_\mathrm{delay})$ given in equation (\ref{Tdiff}):
\begin{eqnarray}
\label{Tdiff}  && t_\mathrm{delay} \equiv \Delta t_f - \Delta t_i
\nonumber\\ && = 9.82 \times 10^{-6} \, \mathrm{s} \left(
\frac{M_\mathrm{dm}}{M_\odot} \right) \ln{ \left|
\frac{x_\perp^2}{(x_m + v_m t_\mathrm{obs})^2 + b^2} \right|} \nonumber\\
&& = -9.82 \times 10^{-6} \, \mathrm{s} \left(
\frac{M_\mathrm{dm}}{M_\odot} \right) \ln{ \left| 1 +
\frac{2 x_m v_m t_\mathrm{obs} + v_m^2 t_\mathrm{obs}^2}{x_\perp^2} \right|}
\mathrm{.}
\end{eqnarray}

If this physical delay were an entirely observable effect, one would note immediately its linear dependence on mass $(M_\mathrm{dm})$ and its logarithmic dependence on position, $x_\perp$.  This would result in objects such as the Large Magellanic Cloud and Andromeda having much larger physical effects than anything transiting close to the LOS.  The resolution to this problem, discussed further in section 3, is that not all of the physical effect is observable, and it is only the observable component that is of interest.
I also note that the final expression in equation (\ref{Tdiff}) is still completely
analytically correct, but will prove useful when the special case of $v_m t_\mathrm{obs} \ll x_\perp$ is examined below.

\section{Pulsar Timing and Detectability}

The change in the pulse arrival time, however, as given by
equation (\ref{Tdiff}), is the magnitude of the physical signal induced by a transiting clump of matter, not the magnitude of the observable signal.  The reason for this is that, a priori, the period and period derivative of the pulsar are not known.  As such, when one measures the period $(P)$ and the period derivative $(\dot{P})$, one is not measuring the actual physical period and physical spin-up or spin-down of the pulsar.  Instead, there are a number of effects that are confounded with the intrinsic period and period derivative, and they contribute, too, to what is observed for $P$ and $\dot{P}$.

Any signal that contributes linearly with time to the arrival time of pulses will be confounded with the period, and any signal contributing quadratically with time will be confounded with the period derivative.  The parallel component of the velocity of the pulsar relative to Earth is an example of an effect inseparable from the period.  Because there is no way to measure the velocity of the pulsar independently, as the pulsar moves along the LOS, the distance along the LOS changes by an amount equal to the pulsar's velocity $(v_p)$ times the pulsar's period $(P)$, causing a change in the apparent period of $v_p P / c$, where $c$ is the speed of light.  Likewise, the pulsar's parallel acceleration causes an analogous shift in the apparent period derivative, $\dot{P}$.

The way that pulsar astronomy accounts for these inseparable effects is to subtract them out.  This is accomplished through the measurement of the pulse arrival times of many pulses over considerably long temporal baselines, and constructing the best linear and quadratic fits to the pulse arrival times.  From the linear fit and the uncertainty in it, the period $(P)$ and the period uncertainty $(\Delta P)$ of the pulsar are determined.  From the quadratic fit and its uncertainty, the period derivative $(\dot{P})$ and its uncertainty $(\Delta \dot{P})$ are determined.

What this means for transiting clumps of matter is that their contribution to the change in pulse arrival time must be decomposed into a best fit linear $+$ quadratic piece, which will be absorbed into the period, period derivative, and their uncertainties.  The remaining timing residuals are the detectable portions of the physical effect detailed in equation (\ref{Tdiff}).  It becomes immediately clear that high-mass objects significantly far away from the LOS will have a significant impact on the observed period, but will have a negligible impact on the timing residuals.  For objects significantly far away, i.e., $x_\perp \gg v_m t_\mathrm{obs}$, it becomes useful to use the Mercator expansion of the natural logarithm in equation (\ref{Tdiff}), obtaining
\begin{eqnarray}
\label{MercExp}
t_\mathrm{delay} &\simeq& 9.82 \times 10^{-6} \, \mathrm{s} \left(
\frac{M_\mathrm{dm}}{M_\odot} \right) \big[ \frac{2 x_m v_m t_\mathrm{obs}}{x_\perp^2} \nonumber\\ &-& \frac{2 x_m^2 v_m^2 t_\mathrm{obs}^2}{x_\perp^4} + \frac{8 x_m^3 v_m^3 t_\mathrm{obs}^3}{3 x_\perp^6} - \mathrm{...} \big] \mathrm{,}
\end{eqnarray}
where the first term in the series will alter the period, the second will alter the period derivative, and only the third and higher terms will affect the timing residual.  The effect on the timing residual, i.e., the detectable part, scales as $(v_m t_\mathrm{obs} / x_\perp)^3$, for objects far enough away that $x_\perp \gg v_m t_\mathrm{obs}$ is a valid approximation.  As $(v_m t_\mathrm{obs} / x_\perp)^3$ is a number of order $\sim 10^{-24}$ for the Large Magellanic Cloud, it is therefore completely negligible.

\begin{figure}
\includegraphics[width=3.5in]{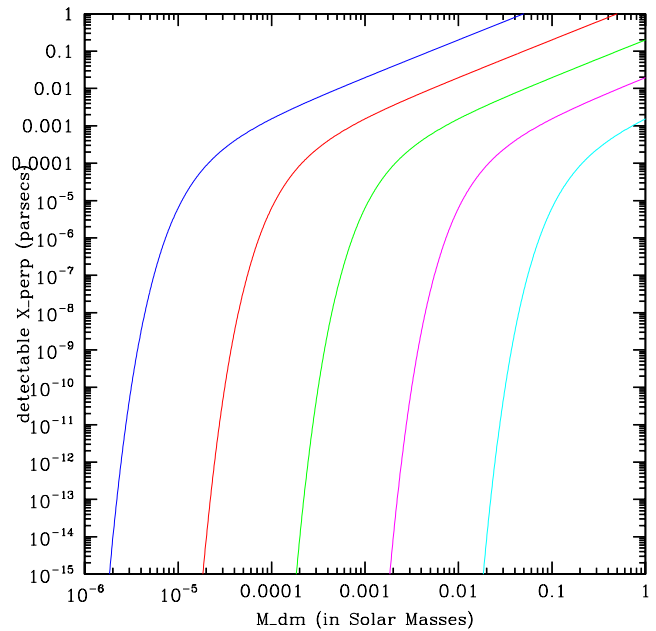}
\caption{The distance within which a clump of matter of mass $M_\mathrm{dm}$ must come of the line-of-sight in order to be detectable in the timing residual.  The five curves (left to right) are for pulsars that can deliver timing accuracies of $1 \,$ns, $10 \,$ns, $100 \,$ns, $1 \, \mu$s, and $10 \, \mu$s, respectively.  This is for a relatively conservative estimate of $v_m t_\mathrm{obs} \simeq 10^{-3} \, \mathrm{pc}$.}
\label{xvsm}
\end{figure}

Since I have demonstrated that the effects of matter at too great a distance will be negligible in its effects on pulsar timing data, I am compelled to examine matter relatively close to the LOS.  This begs the question "how close must one be?" to the LOS to have a shot at detecting a clump of mass $M_\mathrm{dm}$.  I approximate the answer by assuming the detectable component of the time delay to be the entire time delay of equation (\ref{Tdiff}) with the first two terms of equation (\ref{MercExp}) subtracted out.  I then numerically invert the equation to solve for the necessary $x_\perp$ (which will be equivalent to the necessary impact parameter, $b$) to be able to detect the matter clump in the timing residual.  It is assumed that the transiting clump of matter will pass through the point of closest approach during the timeframe of observation.  (The point of closest approach is also known as ``conjunction'' in some circles.)  Figure \ref{xvsm} shows the maximum possible $x_\perp$ in parsecs that will result in a detectable signature in the timing residual as a function of mass $(M_\mathrm{dm})$ of the matter and timing accuracy $(t_\mathrm{res})$ of the pulsar.

Based on this derived distance, henceforth referred to as $b_\mathrm{max}$, or the maximal impact parameter that will allow for detection of transiting matter, a number of interesting quantities can be derived.  Again, assuming the clump of matter of interest has a component of its velocity $v_m$, where $v_m$ is in the plane perpendicular to the LOS, but is also orthogonal to the direction of the impact parameter, $b$.  I assume that the pulsar is observed over a timescale $t_\mathrm{obs}$.  In this case, there are four regions of interest to consider, as to whether the matter clump will pass within detectable range of the LOS, as illustrated in figure \ref{intersect} below.  For a clump originating in the innermost region containing point 1, the probability of detection is unity, whereas in the outermost region (containing point 4), the probability is zero.  In the region containing point 2, at a radius $b_\mathrm{max} < r < v_m t_\mathrm{obs}$, and the region containing point 3, at a radius $v_m t_\mathrm{obs} < r < v_m t_\mathrm{obs} + b_\mathrm{max}$, the probabilities $(p)$ are given by using the appropriate trigonometric laws in equation (\ref{probs}), assuming that the orientation of $v_m$ is completely random.
\begin{eqnarray}
\label{probs}
p&=&1 \qquad \qquad \qquad \qquad \qquad \qquad r < b_\mathrm{max} \mathrm{,} \nonumber\\
p&=& \frac{1}{\pi} \arcsin{\frac{b_\mathrm{max}}{r}} \qquad \qquad \qquad b_\mathrm{max} < r < v_m t_\mathrm{obs} \mathrm{,} \nonumber\\
p&=& \frac{1}{\pi} \arccos{\frac{r^2+v_m^2 t_\mathrm{obs}^2 - b_\mathrm{max}^2}{2 r v_m t_\mathrm{obs}}} \, \, v_m t_\mathrm{obs} < r < v_m t_\mathrm{obs} + b_\mathrm{max} \mathrm{,} \nonumber\\
p&=& 0, \qquad \qquad \qquad \qquad \qquad \qquad r > v_m t_\mathrm{obs} + b_\mathrm{max} \mathrm{.}
\end{eqnarray}
Also note that equation (\ref{probs}) assumes that $v_m t_\mathrm{obs} > b_\mathrm{max}$; if this is not true, then there is no region containing point 2, and the quoted probability for the region containing point three is valid for the entire range $b_\mathrm{max} < r < v_m t_\mathrm{obs} + b_\mathrm{max}$.

\begin{figure}
\includegraphics[width=3.5in]{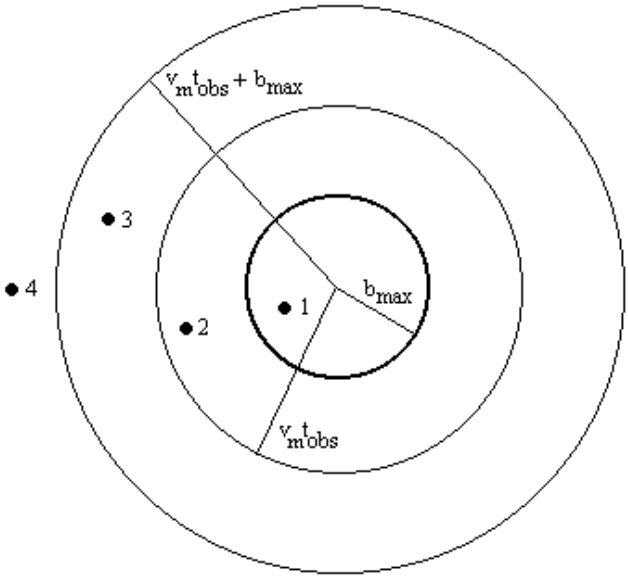}
\caption{Face on view of the LOS to a pulsar.  Any clump of matter of mass $M_\mathrm{dm}$ that comes within a distance $b_\mathrm{max}$, as given by figure \ref{xvsm}, will be detectable.  The probability of a clump of matter coming within a distance $b_\mathrm{max}$ of the LOS for each of the four possible regions illustrated here are calculated in equation (\ref{probs}).}
\label{intersect}
\end{figure}

For each pulsar that is observed, the probability of observing a clump of matter passing within the critical distance to the LOS can be calculated.  Every pulsar will have its own properties, including its own timing accuracy $(t_\mathrm{res})$, period $(P)$, period derivative $(\dot{P})$, uncertainties in the period $(\Delta P)$ and period derivative $(\Delta \dot{P})$, and its distance from Earth $(l)$.  This probability, in addition to its dependence on $t_\mathrm{obs}$ and $l$, will also depend on how long the pulsar is observed for $(t_\mathrm{obs})$, what the mass of the matter clump in question is $(M_\mathrm{dm})$, the velocity of the transiting matter in the plane perpendicular to the LOS $(v_m)$, what fraction of the galaxy (by mass) is locked up in clumps of mass of $M_\mathrm{dm}$ $(f_\mathrm{dm})$, as well as the total mass of the galaxy, $M_\mathrm{gal}$.

The probability is calculated as follows: I assume an NFW profile for the matter in our galaxy \citep{NFW:1997}, with a turnover radius at 25 kpc and our local position located at 8 kpc, both measured from the galactic center.  I further assume that a fraction $f_\mathrm{dm}$ of the matter is locked away in clumps of mass $M_\mathrm{dm}$, which move randomly through the galaxy in all possible directions with a typical rms velocity of $200 \, \mathrm{km} \, \mathrm{s}^{-1}$.  This yields an rms velocity in the plane perpendicular to the LOS of $141 \, \mathrm{km} \, \mathrm{s}^{-1}$.  All volume in a cylindrical shape of length $l$ and radius $b_\mathrm{max}$ is sensitive to transiting matter clumps.

\begin{figure}
\includegraphics[width=3.5in]{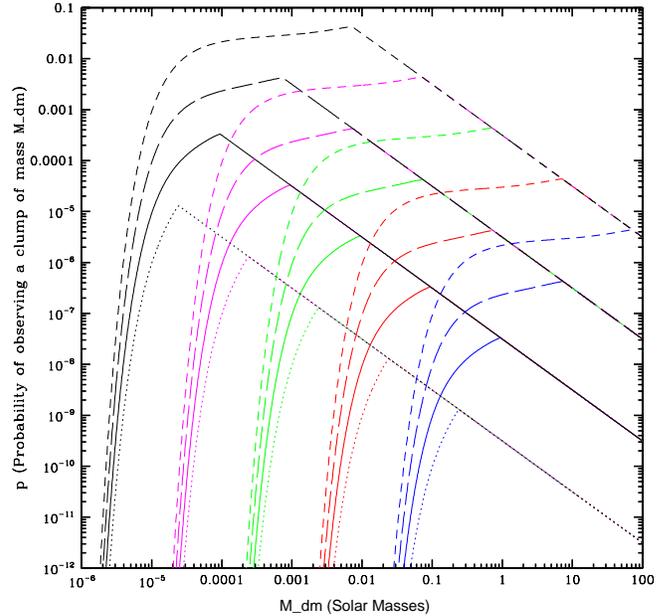}
\caption{Probability of detecting transiting matter of mass $M_\mathrm{dm}$ by looking at one given pulsar with timing accuracy $t_\mathrm{res}$ for an observing time $t_\mathrm{obs}$.  Short-dashed lines are for an observing time of 1000 years, long-dashed for 100 years, solid for 10 years, and dotted for 1 year.  Lower masses can be probed with higher probabilities by improving the timing accuracy; the leftmost set of curves corresponds to a timing accuracy of $1 \, $ns, and moving right, the following sets correspond to accuracies of $10 \, $ns, $100 \,$ns, $1 \, \mu$s, and $10 \, \mu$s.  Note that at sufficiently high masses, improvements in the timing accuracy offer no increase in probability. }
\label{PvsM}
\end{figure}

Figure \ref{PvsM} illustrates the probability of observing one or more clumps of transiting matter as a function of $M_\mathrm{dm}$, $t_\mathrm{obs}$, and $t_\mathrm{res}$.  $M_\mathrm{dm}$ is varied continuously, and the probability $(p)$ is shown for five different timing residuals of $1 \, $ns, $10 \,$ns, $100 \,$ns, $1 \, \mu$s, and $10 \, \mu$s are illustrated, along with four different observing times of 1, 10, 100, and 1000 pulsar-years.  $l$ is fixed at $1 \,$kpc, but the probability is directly proportional to $l$.  The probability $p$ is also directly proportional to $f_\mathrm{dm}$ (assumed to be 1) and $M_\mathrm{gal}$ (assumed to be $1.2 \times 10^{12} M_\odot$), and depends on $v_m$ (assumed to be $141 \, \mathrm{km} \, \mathrm{s}^{-1}$) in the same way as it depends on $t_\mathrm{obs}$ (as shown in figure \ref{PvsM}).  These probabilities are small and highly dependent on the above assumptions, but there may be ways to bring the total probability of detecting such matter close to unity, which I elaborate on in section 5.

\section{Signal Templates and Fitting}

\begin{figure}
\includegraphics[width=3.5in]{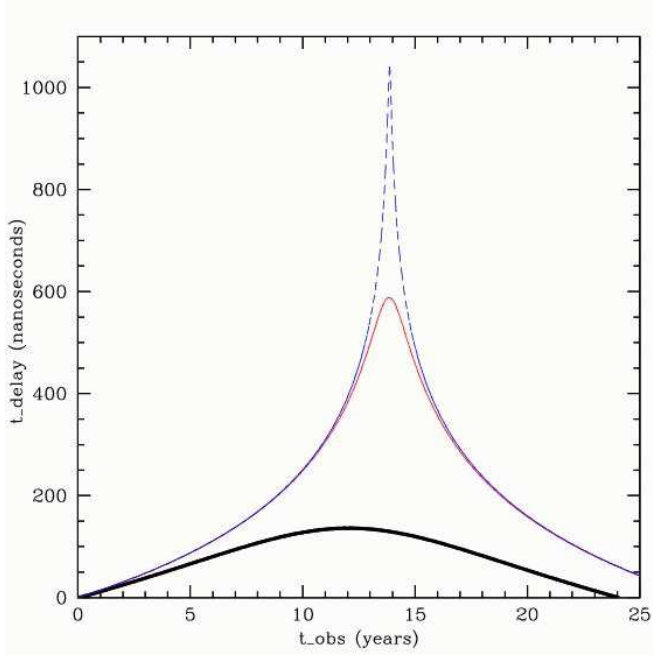}
\caption{The physical time delay induced by a transiting clump of matter of mass $M_\mathrm{dm} = 0.01 \, M_\odot$ with an impact parameter of $10^{-3} \, \mathrm{pc}$ (thick solid line), $10^{-4} \, \mathrm{pc}$ (thin solid line), or $10^{-5} \, \mathrm{pc}$ (thin dashed line), velocity $v_m = 141 \, \mathrm{km} \, \mathrm{s}^{-1}$, initial position $|\vec{x}| = 0.002 \, \mathrm{pc}$, and observed for 25 years.  Clumps that pass even closer to the LOS will have greater physical signals, but are extraordinarily infrequent.  Clumps that move with a greater $v_m$ will have their timescales (along the x-axis) compressed, while those that are slower moving will have their timescales stretched.  The signal strength is directly proportional to $M_\mathrm{dm}$.}
\label{physical}
\end{figure}

More important than a calculation of the physical time delay induced by transiting matter, or even than the magnitude of the detectable component, it is vital to uncover what the signature of transiting matter will look like in the pulsar timing data.  As the arriving pulses are very regular, it becomes very easy to calculate the physical signals in the timing residuals induced by the transiting matter, illustrated for a variety of impact parameters in figure \ref{physical} for a clump of mass $M_\mathrm{dm} \simeq 10^{-2} M_\odot$ with a transverse velocity $v_m = 141 \, \mathrm{km} \, \mathrm{s}^{-1}$, 
an initial position $|\vec{x_\perp}| = 0.002 \, \mathrm{pc}$, and an observing time $(t_\mathrm{obs})$ of 25 years.

As stated earlier, however, a signal like that shown in figure \ref{physical} will never show up as illustrated in the timing residuals.  What is feasible, however, is to subtract off the best linear and quadratic fit, and extract the signal arising from that.  One key point is that {\it the best quadratic and linear fits to the data will change over time as more data is collected}.  Therefore, I have two possibilities at my disposal for searching for transiting matter with pulsars: one is to match the theoretical signal templates with the observed timing residuals, and the other is to look for systematic changes with observing time of the period and period derivative.  I detail both methods here.  

I perform a least-squares fit for a second degree polynomial in $t_\mathrm{obs}$ for the first halves of each of the curves (between 12 and 14 years of observing time) as well as for the full 25 years, calculating the linear and quadratic fits as well as the templates for the remaining signal that will be observable in the timing residual data.  The three curves from figure \ref{physical} are analyzed as follows: the physical time delay data is fitted to a second-degree polynomial using least-squares.  The fit for the first half of the curve (up until the matter makes its closest approach to the LOS) is calculated first.  Subsequently, the fit to the entire 25 year simulation is calculated.  The differences between the least-squares fit and the physical delay results in the signal that will show up in the timing residuals; all of these are illustrated in figure \ref{templates} below.  It is immediately clear that observing the complete transit, rather than just a partial component of it, results in a far greater magnitude of observable signal.  Additionally, smaller impact parameters result in a greater signal strength.

As can also be seen from comparing the graphs on the left-hand side of figure \ref{templates} with those on the right-hand side, the least-squares fits change dramatically from the partial transit to the complete transit, as well as producing a greater observable signal in the residuals.  However, the differences in the polynomial fits correspond to a difference in the inferred period $(P)$ and period derivative $(\dot{P})$.  The differences in both $P$ and $\dot{P}$ are comparable to the inherent uncertainties in $P$ and $\dot{P}$ for many millisecond pulsars.  For the case of an impact parameter of $10^{-3} \,$pc, I calculate a perceived change in $P$ of $4.1 \times 10^{-18} \,$s and a change in $\dot{P}$ of $3.8 \times 10^{-26}$, where I remind the reader that $\dot{P}$ is dimensionless.  For an impact parameter of $10^{-4} \,$pc, $P$ changes by $2.0 \times 10^{-16} \,$s and $\dot{P}$ changes by $5.3 \times 10^{-25}$.  Finally, for an impact parameter of $10^{-5} \,$pc, $P$ changes by $2.7 \times 10^{-16} \,$s and $\dot{P}$ changes by $7.0 \times 10^{-25}$. Changes in $P$ and/or $\dot{P}$ over very long timescales of this magnitude may therefore be indicative of transiting clumps of matter.

\begin{figure}
\begin{center}$
\begin{array}{cc}
\includegraphics[width=1.7in]{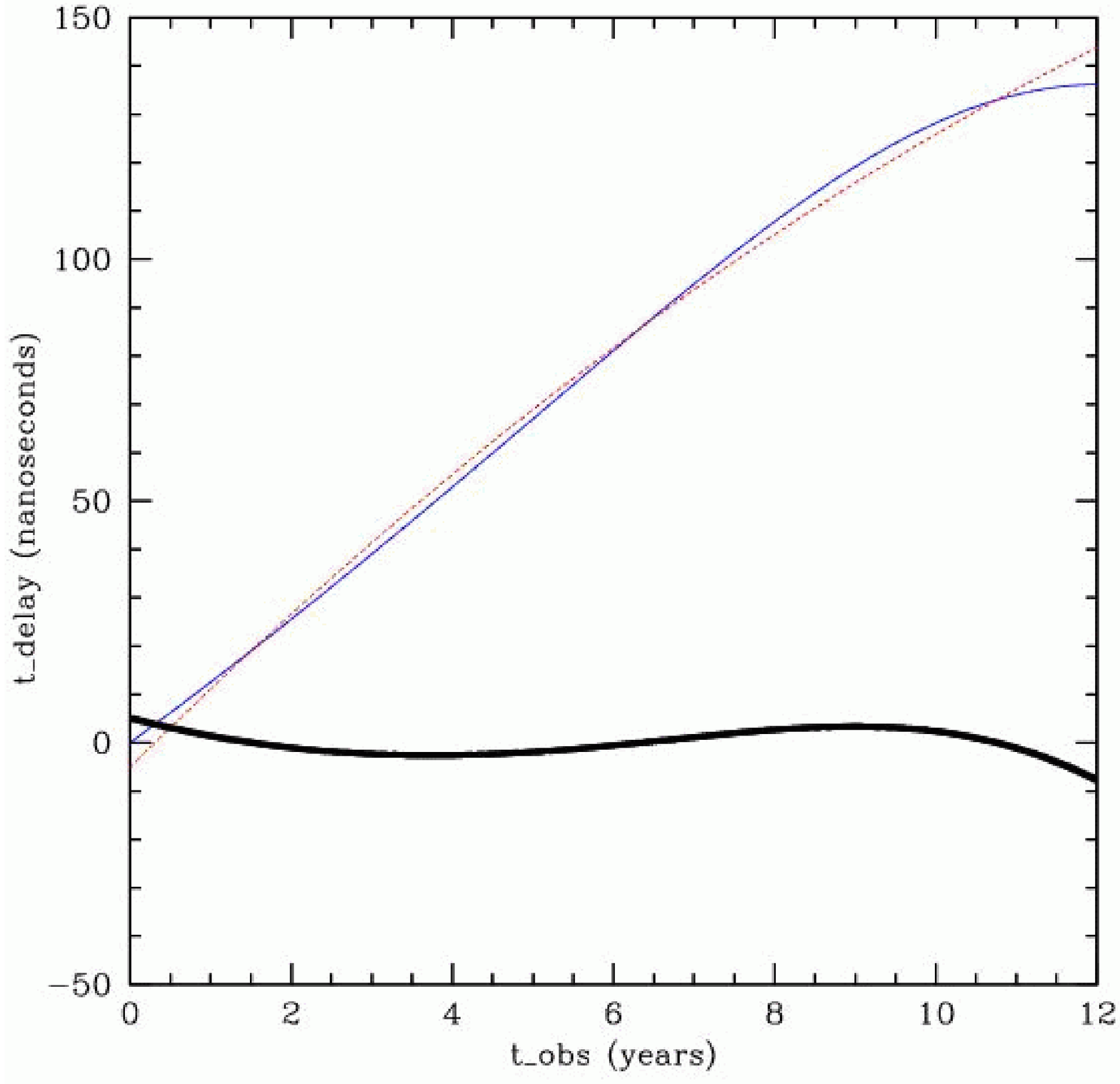} &
\includegraphics[width=1.7in]{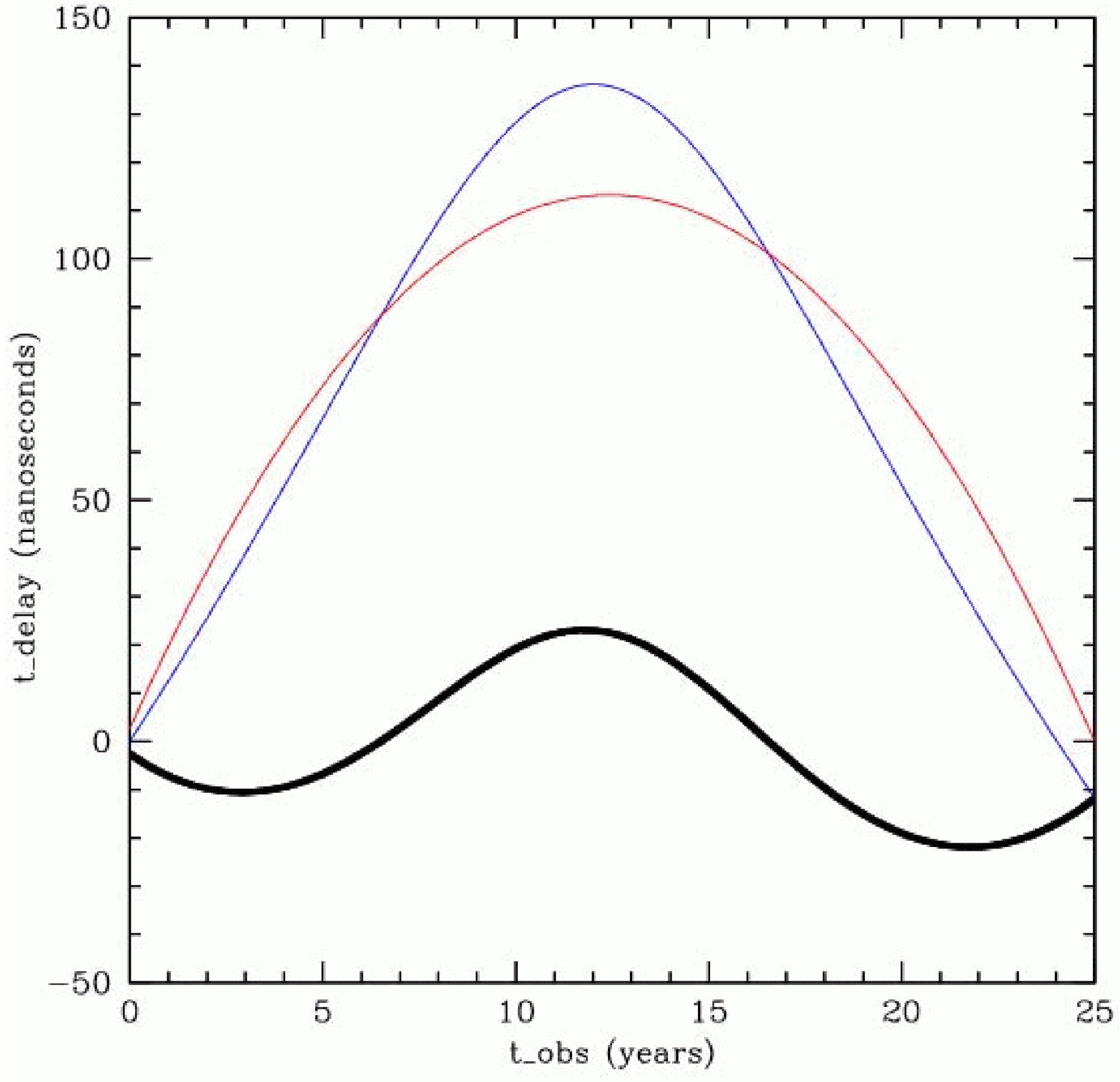} \\
\includegraphics[width=1.7in]{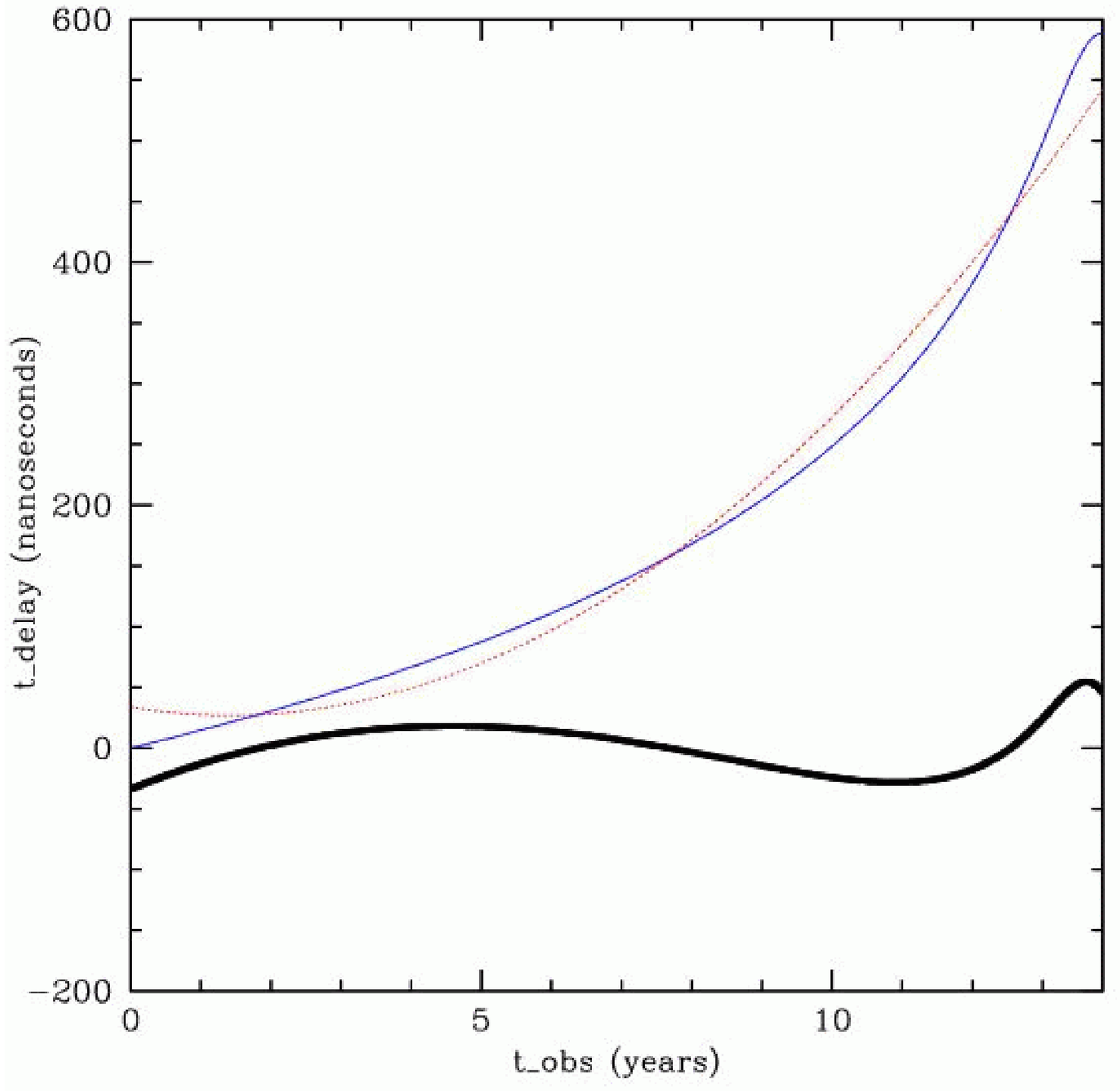} &
\includegraphics[width=1.7in]{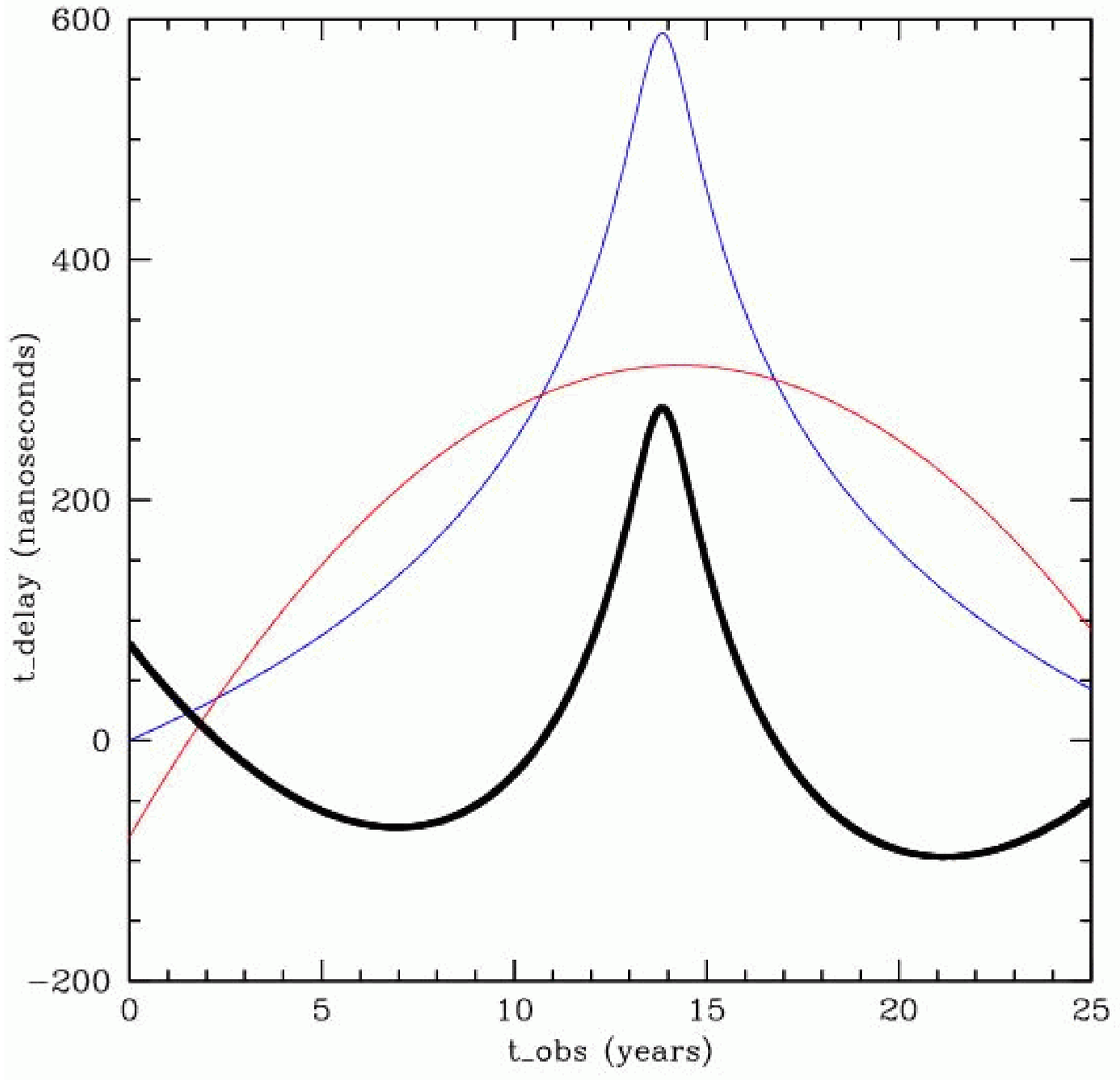} \\
\includegraphics[width=1.7in]{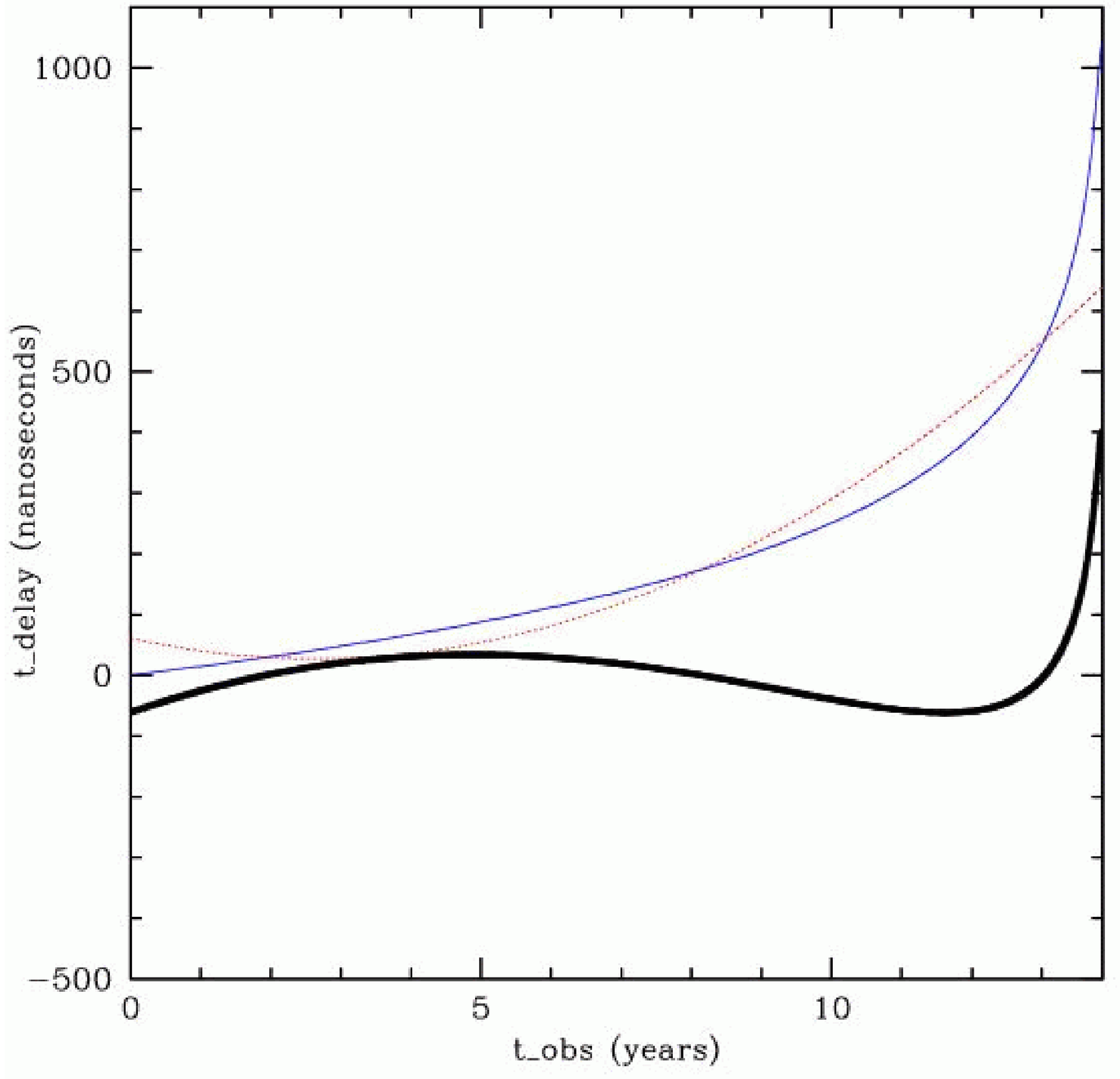} &
\includegraphics[width=1.7in]{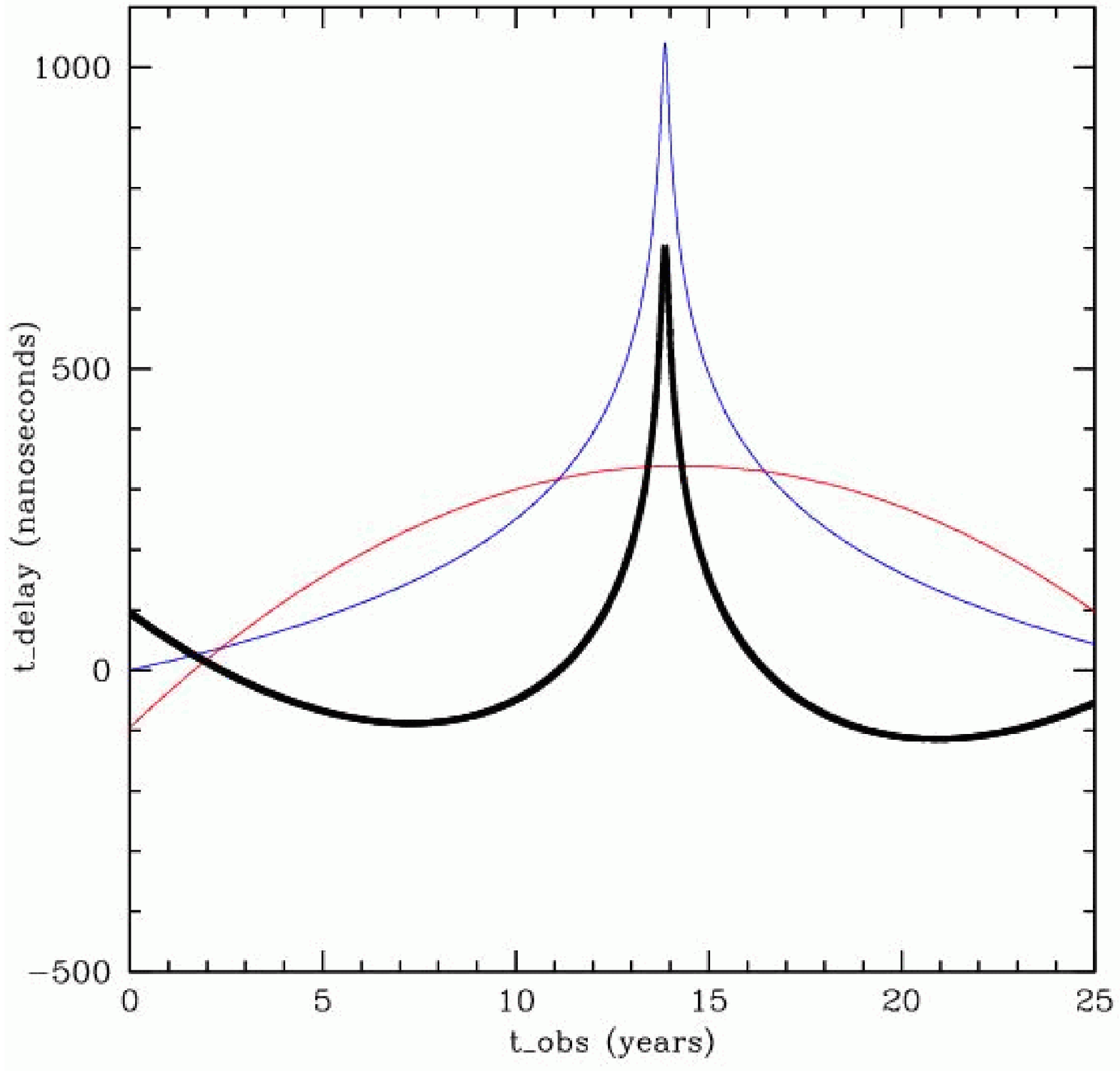} 
\end{array}$
\end{center}
\caption{Templates for the signal that will show up in the pulsar timing residual data if there is a transiting clump of matter close to the LOS.  The upper two figures are for an impact parameter of $10^{-3} \,$pc, the middle two figures are for an impact parameter of $10^{-4} \,$pc, and the lower two figures illustrate an impact parameter of $10^{-5} \,$pc.  The figures on the left demonstrate motion of the clump.  The thin solid lines represent the physical time delay, the dotted lines represent the best second-order polynomial least-squares fit, and the thick solid line, the difference of the two, represents the signal that should be present in the pulsar timing data.  This is for a transiting mass $M_\mathrm{dm} = 0.01 \, M_\odot$ with velocity $v_m = 141 \, \mathrm{km} \, \mathrm{s}^{-1}$; the magnitude of the signal scales directly proportional to mass, and the timescale along the x-axis scales proportionate to velocity. }
\label{templates}
\end{figure}

\section{Discussion}

I have demonstrated the quantifiable, observable effects of transiting matter on pulsar timing measurements.  Perhaps most importantly, I have crafted templates detailing the observable signal in the timing residuals as well as predicting the corresponding changes in $P$ and $\dot{P}$ which accompany such a transit.  These are effects which can lead to the indirect detection of either collapsed baryonic matter or dark matter in between Earth and a millisecond pulsar.

Current timing accuracies for millisecond pulsars are typically of order $\sim 1 \, \mu$s, although there do exist a number of millisecond pulsars with accuracies of $\sim 100 \, $ns \citep{Kaspi:94,vanStraten:01,Jacoby:05,Hotan:2006}, such as B1937+21, J1909-3744, J0437-4715, and J1713+0747.  Additionally, uncertainties in $P$ are typically $\mathcal{O}(10^{-16} \, \mathrm{s})$ and uncertainties in $\dot{P}$ are typically $\mathcal{O}(10^{-25})$, although some are more accurate, for example, J1909-3744 has an uncertainty in $P$ of $2.0 \times 10^{-18} \, \mathrm{s}$ and in $\dot{P}$ of $2.0 \times 10^{-25}$ \citep{ATNF}.  While timing accuracies will have to improve by two orders of magnitude to detect clumps of mass $M_\mathrm{dm} \simeq 10^{-4} \, M_\odot$ with somewhat fortuitous transit parameters, measurements of the uncertainties in $P$ are nearly at the required level to give evidence of a transit today.  (I remind the reader that the $M_\mathrm{dm}$ used in earlier sections are $0.01 \, M_\odot$.)  I recommend that everyone working with millisecond pulsars compare their timing residuals to these templates and also look for the corresponding systematic, coherent changes over time in the period and period derivatives of their pulsars.

While the probability of observing a transit sufficiently close to the LOS in observing any one pulsar is exceedingly small, there are four reasons for optimism.  First, the analysis presented is for one individual pulsar, but over 100 millisecond pulsars are presently known, with the total detectable galactic pulsar population estimated to be approximately 30,000 \citep{Kiel:2007}, many of which will be discovered and timed to unprecedented accuracy with the Square Kilometer Array \citep{SKA:2005}.  Second, this assumes a distance to the pulsar of interest of $1 \, \mathrm{kpc}$.  While this is a typical distance to the most accurate millisecond pulsars at present, millisecond pulsars have been detected as far away as the globular clusters, and new estimates are that millisecond pulsar timing is feasible for objects as far away as $\sim 1.2 \, \mathrm{Mpc}$ \citep{LOFAR:2007}, which would open up an extremely long baseline to millisecond pulsars in other galaxies (e.g., Andromeda), should this be correct.  Third, the gravitational effects should be easily separable from other effects (e.g., dust lanes, hot gas, etc.) because of the greyness of the gravitational effects.  While electromagnetic effects cause shifts of differing magnitudes as a function of frequency \citep{Snez:2007}, the gravitational time delay is frequency independent.  Other gravitational sources, such as gravitational scattering of the pulsar with a star or planet, or a gravitational wave effect on timing, will have a different template signal and can always be discriminated against by continued timing measurements.  Finally, improvements in software \citep{TEMPO:2006} combined with future improvements in instrumental timing accuracies \citep{Shearer:08} may indeed lead to residuals knowable to $\mathcal{O}( 1 \, \mathrm{ns})$, which would be a great boon to the range of detectability of transiting matter.  This method is primarily aimed at pulsars located in the galactic disk, where precision timing is most effective.  It is worth noting that there is discussion as well of precision timing and the possibility of observing the Shapiro effect for pulsars in the halo as well \citep{HaloPulsars:2007}.

Timing of millisecond pulsars is sufficiently accurate that it should be able to determine whether or not matter of a given mass has transited close to the LOS to the pulsar over the timescales of observation.  Projects such as the Parkes Pulsar Timing Array \citep{Manchester:07} are just coming online at present, and have the ability to detect these effects.  While this analysis is not applicable to extended sources that cross the LOS, it is applicable to any source that does not, as well as any point-like source that does.  Future work may attempt to address that problem, made more complex by the changing effective gravitational potential.  Such a template would be extremely useful for searching for various types of dark matter, such as Scalar Miniclusters \citep{Zurek:2007} or WIMPy microhalos with small impact parameters.  


I acknowledge Kathryn Zurek, first and foremost, for her contributions to this work during the early stages, without which this paper would not have been possible.  I also acknowledge Snezana Stanimirovic for useful discussions concerning pulsar astronomy and timing, and Jim Chisholm for
discussions on Primordial Black Holes.  Many thanks as well to Craig Hogan, Jim Fry and Daniel Eisenstein for comments on drafts of this paper.


\begin{thebibliography}{}

\bibitem[Akerib et al.(2005)]{CDMS:2005} Akerib, D.~S., et al.\ 
2005, \prd, 72, 052009 

\bibitem[Alcock et al.(1998)]{Alcock:98} Alcock, C., et al.\ 
1998, \apjl, 499, L9 

\bibitem[Backer et al.(1982)]{MSP:82} Backer, D.~C., Kulkarni, 
S.~R., Heiles, C., Davis, M.~M., \& Goss, W.~M.\ 1982, \nat, 300, 615

\bibitem[Cole et al.(2005)]{Cole:2005sx} Cole, S., et al.\ 2005, 
\mnras, 362, 505

\bibitem[Cordes(2005)]{SKA:2005} Cordes, J.~M.\ 2005, 
Astronomical Society of the Pacific Conference Series, 345, 461 

\bibitem[Fargion \& Conversano(1997)]{Fargion:97} Fargion, D., \& 
Conversano, R.\ 1997, \mnras, 285, 225

\bibitem[Fukugita \& Peebles(2004)]{FP:04} Fukugita, M., \& 
Peebles, P.~J.~E.\ 2004, \apj, 616, 643 

\bibitem[Hobbs et al.(2006)]{TEMPO:2006} Hobbs, G., Edwards, R., 
\& Manchester, R.\ 2006, Chinese Journal of Astronomy and Astrophysics 
Supplement, 6, 189 

\bibitem[Hosokawa et al.(1999)]{Hosokawa:1999} Hosokawa, M., Ohnishi, 
K., \& Fukushima, T.\ 1999, \aap, 351, 393

\bibitem[Hotan et al.(2006)]{Hotan:2006} Hotan, A.~W., Bailes, M., 
\& Ord, S.~M.\ 2006, \mnras, 369, 1502 

\bibitem[Jacoby et al.(2005)]{Jacoby:05} Jacoby, B.~A., Hotan, 
A., Bailes, M., Ord, S., \& Kulkarni, S.~R.\ 2005, \apjl, 629, L113

\bibitem[Kaspi et al.(1994)]{Kaspi:94} Kaspi, V.~M., Taylor, 
J.~H., \& Ryba, M.~F.\ 1994, \apj, 428, 713 

\bibitem[Kiel et al.(2007)]{Kiel:2007} Kiel, P., Hurley, J., 
Bailes, M., \& Murray, J.\ 2007, ArXiv e-prints, 711, arXiv:0711.4138 

\bibitem[Konacki \& Wolszczan(2003)]{Konacki:2003} Konacki, M., \& 
Wolszczan, A.\ 2003, \apjl, 591, L147

\bibitem[Larchenkova \& Doroshenko(1995)]{LD:95} Larchenkova, 
T.~I., \& Doroshenko, O.~V.\ 1995, \aap, 297, 607

\bibitem[Larchenkova \& Lutovinov(2007)]{HaloPulsars:2007} Larchenkova, 
T.~I., \& Lutovinov, A.~A.\ 2007, Astronomy Letters, 33, 455 

\bibitem[Manchester et al.(2005)]{ATNF} Manchester, R.~N., 
Hobbs, G.~B., Teoh, A., \& Hobbs, M.\ 2005, \aj, 129, 1993 

\bibitem[Manchester(2007)]{Manchester:07} Manchester, R.~N.\ 2007, 
ArXiv e-prints, 710, arXiv:0710.5026

\bibitem[Navarro et al.(1997)]{NFW:1997} Navarro, J.~F., Frenk, 
C.~S., \& White, S.~D.~M.\ 1997, \apj, 490, 493 

\bibitem[Seto \& Cooray(2007)]{Seto:2007} Seto, N., \& Cooray, 
A.\ 2007, \apjl, 659, L33 

\bibitem[Shapiro(1964)]{Shapiro:1964} Shapiro, I.~I.\ 1964, Physical 
Review Letters, 13, 789 

\bibitem[Shearer(2008)]{Shearer:08} Shearer, A.\ 2008, ArXiv 
e-prints, 801, arXiv:0801.0314 

\bibitem[Siegel et al.(2007)]{Siegel:2007fz} Siegel, E.~R., 
Hertzberg, M.~P., \& Fry, J.~N.\ 2007, \mnras, 382, 879

\bibitem[van Leeuwen \& Stappers(2007)]{LOFAR:2007} van Leeuwen, 
J., \& Stappers, B.\ 2007, ArXiv e-prints, 710, arXiv:0710.0675 

\bibitem[van Straten et al.(2001)]{vanStraten:01} van Straten, W., 
Bailes, M., Britton, M., Kulkarni, S.~R., Anderson, S.~B., Manchester, 
R.~N., \& Sarkissian, J.\ 2001, \nat, 412, 158 

\bibitem[Weisberg \& Stanimirovi{\'c}(2007)]{Snez:2007} Weisberg, 
J.~M., \& Stanimirovi{\'c}, S.\ 2007, SINS - Small Ionized and Neutral 
Structures in the Diffuse Interstellar Medium, 365, 28 

\bibitem[Wilkinson \& Evans(1999)]{Wilk:1999} Wilkinson, M.~I., 
\& Evans, N.~W.\ 1999, \mnras, 310, 645 

\bibitem[Will(1980)]{Will:80} Will, C.~M.\ 1980, Ninth Texas 
Symposium on Relativistic Astrophysics, 336, 307

\bibitem[Zurek et al.(2007)]{Zurek:2007} Zurek, K.~M., Hogan, 
C.~J., \& Quinn, T.~R.\ 2007, \prd, 75, 043511 

\end{thebibliography}

\end{document}